\documentclass[manuscript]{aastex}

\begin{document}

\title{Observations of a possible new soft gamma repeater, SGR1801-23}

\author{T. Cline\altaffilmark{1}, D. D. Frederiks\altaffilmark{2},
S. Golenetskii\altaffilmark{2}, K. Hurley\altaffilmark{3}
C. Kouveliotou\altaffilmark{4}, E. Mazets\altaffilmark{2}, and
J. van Paradijs\altaffilmark{5,6}}

\altaffiltext{1}{NASA Goddard Space Flight Center, Greenbelt, MD 20771}
\altaffiltext{2}{Ioffe Physical-Technical Institute, St. Petersburg, 194021, Russia}
\altaffiltext{3}{University of California, Berkeley, Space Sciences Laboratory, Berkeley, CA 94720-7450, khurley@sunspot.ssl.berkeley.edu}
\altaffiltext{4}{Universities Space Research Association at NASA Marshall Space Flight Center, ES-84, Huntsville AL 35812}
\altaffiltext{5}{Astronomical Institute `Anton Pannekoek',
University of Amsterdam, The Netherlands}
\altaffiltext{6}{Physics Department, University of Alabama in Huntsville, Huntsville, AL 35899}

\begin{abstract}

We report on two observations of a soft bursting source in 1997 June, whose
time histories and energy spectra are consistent with those of the soft gamma repeaters.  
The source can only be localized to an $\approx$ 3.8$\rm ^o$ long error box 
in the direction of the Galactic center, whose
area is $\approx$ 80 arcmin$^2$.  The
location of the source, while not consistent with that of any of the four
known soft repeaters, is consistent with those of several known and possible
supernova remnants.

\end{abstract}


\section{Introduction}

Soft gamma repeaters (SGRs) are neutron stars in or near radio or optical supernova remnants.  
There is good evidence that they are 'magnetars', i.e. neutron stars in which the magnetic field energy dominates all other sources of energy, including rotation (Duncan \& Thompson 1992).  In the case of SGR1806-20, evidence for this comes from observations of the period and period derivative of the quiescent soft X-ray emission (Kouveliotou et al. 1998).  In
the case of SGR1900+14, evidence comes from observations of both the spindown and
of a giant flare (Kouveliotou et al. 1999, Hurley et al. 1999a; however, see Marsden, Rothschild,
\& Lingenfelter 1999 for a different interpretation).  The magnetar model (Thompson and Duncan
 1995) predicts
a galactic birth rate of $\approx$ 1-10/10000 y, and a lifetime of $\approx$ 10000 y, so at any
given time, up to 10 magnetars could be active.  This is consistent with observational
estimates of the magnetar birth rate and of the total number in the Galaxy (Kouveliotou
et al. 1998).  Only four have been identified to date, however,
and various studies have placed upper limits on the number of active SGRs (e.g., Norris et al.
1991, Kouveliotou et al. 1992, Hurley et al. 1994).  Taking the galactic magnetar census
is therefore an interesting exercise for understanding the formation and
life cycles of these unusual
objects.

In 1997 June, during a period when SGR1806-20 was undergoing a phase of intense
activity, two bursts were observed whose positions were close to, but clearly 
inconsistent with that of this source.  It was hoped that this new source would
remain active, allowing a better determination of its position, but to date this
has not happened.  Therefore we present the existing data at this time, even though
the picture is still incomplete.

\section{Observations}

The two bursts were observed by four instruments: BATSE - CGRO (Meegan et al. 1996), 
Konus-A aboard the Kosmos spacecraft (Aptekar et al. 1997), 
Konus-W aboard the Wind spacecraft (Aptekar et al. 1995), and the GRB
experiment aboard \it Ulysses \rm (Hurley et al. 1992). Table 1 gives the details of the observations, 
including the time resolutions $\rm \Delta T$ with which each instrument observed
the bursts; the time histories are shown in Figures 1 and 2.  Both are short,
and have soft energy spectra, e.g. consistent with an optically thin
thermal bremsstrahlung (OTTB) function with a kT of $\approx$ 25 keV.  The peak fluxes and fluences are reported in Tables 1 and 2.  Note that the peak flux of the second burst implies that the source is super-Eddington
for any distance $\rm \gtrsim$ 250 pc; at the distance of the Galactic center (see
below) it would be $\rm \gtrsim 1200 L_E$.  All these characteristics are typical of SGRs in general.  In addition, there is evidence in the KONUS-W data for spectral evolution in the
second burst (Frederiks et al. 1998): the initial phase has a spectrum consistent with
an OTTB function with kT $\approx$ 20 keV, softening to kT $\approx$ 9 keV in the final
phase.    

\section{Localization}

The second event was observed by three instruments in high time resolution
modes (Table 1), leading to two statistically independent, narrow triangulation annuli.  
However, since two of the spacecraft
(Konus-W and CGRO) were separated by only 1.4 light-seconds, these annuli have
practically identical centers and radii, and therefore intersect at grazing
incidence to define two long, narrow error boxes, whose lengths are constrained
by the third (Konus-W/BATSE) annulus.  Only one is consistent with 
the BATSE error circle (radius $\approx$ 5$\rm ^o$), but the error box is fully contained within it, and
is therefore not constrained by it.  

The first event
was observed with high time resolution by \it Ulysses \rm, but with time
resolution greater than the event duration by the two Konus instruments, leading to
relatively wide triangulation annuli.  These are consistent with the first
error box, but because this event occurred only $\approx$ 9000 s before the 
second one, the
\it Ulysses \rm-Earth vector moved only slightly between the two, resulting again in 
annuli which intersect the first error box at grazing incidence.  This intersection
is consistent with the coarse localization capabilties of Konus-A and Konus-W.
Table 3 gives the details of the triangulation annuli, and Table 4 gives
the coordinates of the error box.

Initially, it was thought, based on preliminary data, that
a third burst originated from this source on 1997 September 12 (Hurley et al. 1997;
Kouveliotou et al. 1997) and that 
the \it Rossi \rm X-Ray Timing Explorer had observed it in the
collimated field of view of the All-Sky Monitor, providing an error box which 
intersected the annuli (Smith et al. 1997).
However, on this day, the \it Ulysses \rm-Earth
vector was equidistant from this error box and the position of SGR1806-20; thus the triangulation
annulus for either one of these sources would automatically pass very close
to the other.  When the final data were obtained and a more precise annulus
could be obtained, it proved to be consistent with the position of SGR1806-20
to better than 10 $\arcsec$, making this SGR the likely source of this event.
Moreover, it turned out that the burst had entered the RXTE ASM proportional counters
through their sides, and no location information could in fact be extracted from the
data (D. Smith, private communication).  Thus the only information on the location
of this new SGR comes from the triangulation annuli and the BATSE error
circle.

The error box, which is in the direction of the Galactic center, is shown in Figure 3.  
The triangulation annuli of the two bursts may also be combined using the statistical method 
of Hurley et al. (1999b) to derive an error ellipse.  The method gives
an acceptable $\chi^{2}$, but results in an ellipse which is somewhat longer
than the error box and only slightly smaller in area.  Given the density of
possible counterpart sources in the region of Figure 3, the error box is probably
the more useful description of the SGR location.  It 
lies $\rm \approx 0.93^o$ from the position of SGR1806-20.  A timing
error of $\approx$ 39 s would have to be invoked for one spacecraft in each of the 
two observations
to achieve consistency with this SGR, and there is no evidence in any of the data
for such an error.

\section{Discussion}

As the four known SGRs are associated with SNRs, we have searched several
catalogs for possible associations.  The results are shown in Figure 3.  G5.4-1.2,
G6.4-0.1, and G8.7-0.1 (just visible at the left of Figure 3) are from Green (1998).  
G6.0-1.2 is from Goss \& Shaver (1970), and all other
sources are from Reich, Reich, \& F\"{u}rst (1990).  Not all of these
objects are confirmed SNRs.  Of the confirmed SNRs, only G6.4-0.1 (=W28)
is consistent with the error box.
However, this SNR may be associated with the pulsar B1758-23 (Kaspi et al. 1993), which lies outside
the error box.  
G5.4-0.29, G7.2+0.2, and G8.1+0.2 are other possible associations.  
Given that SGR1900+14 lies outside its supernova remnant
(Hurley et al. 1999c), SGR1801-23 could also be associated with
an object such as G5.9-0.4, which lies slightly outside the error box.

The four known SGRs are also quiescent soft X-ray sources (e.g. Hurley et al.
1999d and references therein) with fluxes $\rm \approx 10^{-11} - 10^{-12} erg \,
 cm^{-2}s^{-1}$, i.e. bright enough to be detected not only in pointed
observations, but also in sky surveys.  Accordingly, we have searched the ROSAT catalogs available
through the HEASARC.  Only two objects are close to the error box.  One is
the unidentified source 1WGA J1802.3-2151 in the WGA catalog (White, Giommi, \&
Angelini 1995), which lies slightly outside it.  The other is the diffuse emission
associated with W28.

Finally, it has been suggested that magnetars evolve into anomalous X-ray pulsars (AXPs)
(Kouveliotou et al. 1998).  Sporadic bursts from an AXP could confirm this association.
Accordingly, we have checked the positions of the
six known (Gotthelf \& Vasisht 1998 and references therein) and one proposed (Li \&
van den Heuvel 1999) AXPs, but none lies near this source.

Given the shape and location of the error box, it is not unlikely
that it will cross several interesting objects by chance coincidence, and the nature of this source
therefore remains unknown.  Based on the properties of the
two events observed to date, it most closely resembles an SGR.  Indeed,
SGR1900+14 was discovered when it burst just 3 times in 3 days (Mazets et al. 1979); 13 years
elapsed before it was detected again (Kouveliotou et al. 1993).  Until SGR1801-23
bursts again, allowing a more accurate position to be derived for it, associating
it with an SNR or quiescent soft X-ray source will be difficult.

\acknowledgments
KH is grateful to JPL for \it Ulysses \rm support under Contract 958056,
and to NASA for Compton Gamma-Ray Observatory support under
grant NAG 5-3811.  On the Russian side, this work was supported by
RSA Contract and RFBR grants N97-02-18067 and N99-02-17031.
JvP acknowledges support from NASA grants NAG5-3674 and NAG5-7808.
This study has made use of data obtained from the High Energy Astrophysics 
Science Archive Research Center (HEASARC), provided by NASA's Goddard Space Flight Center.

\newpage

\figcaption{Time history of the first 
burst from SGR1801-23 as observed by \it Ulysses \rm.  
The energy range is $\approx$ 25-150 keV.  The background
rate is indicated by a dashed line.  \label{fig1}}

\figcaption{Time history of the second burst as observed by Konus-W
(14-230 keV), BATSE (25-100 keV), and Ulysses (25-150 keV). The background
rates are indicated by dashed lines. \label{fig2}}

\figcaption{IPN error box for SGR1801-23 (the lines are too
closely spaced to distinguish).  The center is indicated with an asterisk.
Circles give the approximate locations of confirmed and suspected SNRs; the
radii have been taken as half the size given in the catalogs.  Asterisks give
the positions of ROSAT X-ray sources, and two pulsars, PSR1800-21 and B1758-23,
probably associated with SNRs 8.7-0.1 and 6.4-0.1. Coordinates are J2000. \label{fig3}
}

\clearpage
\begin{deluxetable}{cccccc}
\tablecaption{\it IPN Observations of SGR1801-23.}
\tablehead{
\colhead{Date} & \colhead{UT, s} & \colhead{BATSE} & \colhead{Konus-A} &  \colhead{Konus-W} & 
\colhead{\it Ulysses} \\
\colhead{}      & \colhead{}      & \colhead{$\Delta$ T, s} & \colhead{$\Delta$ T, s} & \colhead{$\Delta$ T, s} & \colhead{$\Delta$ T, s}  
}

\startdata
970629 	&	14424 & O\tablenotemark{1}         & 2.0   & 1.472 & 0.03125	\\
970629	&	23492	& TTS\tablenotemark{2}, .064 & O     & 0.002 & 0.03125	\\

\enddata
\tablenotetext{1}{Source was Earth-occulted}
\tablenotetext{2}{Time-to-spill mode: variable time resolution from $\approx$ 5 ms up}

\end{deluxetable}

\clearpage
\begin{deluxetable}{cccc}
\tablecaption{\it Peak fluxes and fluences.}
\tablehead{
\colhead{Date} & \colhead{UT, s} & \colhead{Peak flux, 25-100 keV,} & \colhead{Fluence, 25-100 keV}  \\
\colhead{}      & \colhead{}      & \colhead{over 32 ms} & \colhead{erg cm$^{-2}$ } \\
\colhead{}      & \colhead{}      & \colhead{erg cm$^{-2}$ s$^{-1}$} & \colhead{} 
}

\startdata
970629 	&	14424 & $\rm5 \times 10^{-6}$ & $\rm9 \times 10^{-7}$	\\
970629	&	23492	& $\rm2 \times 10^{-5}$ & $\rm5 \times 10^{-6}$	\\

\enddata
\end{deluxetable}

\clearpage
\begin{deluxetable}{ccccccc}
\tablecaption{\it IPN Localizations of SGR1801-23.  }
\tablehead{
\colhead{} & \colhead{} & \colhead{} & \multicolumn{2}{c}{Annulus center} & \colhead{} & \colhead{} \\
\colhead{Date} & \colhead{UT, s}  & \colhead{Spacecraft} & \colhead{$\rm \alpha(2000)$}& \colhead{$\rm \delta(2000)$} & \colhead{Radius, $\theta$} & \colhead{3$\sigma$ half-width} \\
\colhead{} & \colhead{} & \colhead{} & \colhead{(deg.)} & \colhead{(deg.)} & \colhead{(deg.)}
 & \colhead{(deg.)}
}

\startdata
970629 & 14424 & \it Ulysses \rm - Konus-W & 333.7154 & -25.9376 & 57.2971 & 0.0206	\\
970629 & 14424 & \it Ulysses \rm - Konus-A & 333.6945 & -25.9347 & 57.2813 & 0.0268	\\
970629 & 23492 & \it Ulysses \rm - BATSE   & 333.7050 & -25.9223 & 57.2819 & 0.0030	\\
970629 & 23492 & \it Ulysses \rm - Konus-W & 333.7251 & -25.9253 & 57.2990 & 0.0030  \\
970629 & 23492 & Konus-W - BATSE           & 295.6453 & -15.1844 & 25.1863 & 0.9472  \\

\enddata
\end{deluxetable}

\clearpage
\begin{deluxetable}{ccc}
\tablecaption{\it Triangulation error box of SGR1801-23}
\tablehead{
\colhead{} & \colhead{$\rm \alpha (2000)$, degrees} & \colhead{$\rm \delta(2000)$, degrees}
}
\startdata
Center:     & 270.2454        & -22.9468 \\
Corners:    & 269.6792        & -24.6820 \\
           & 270.8738        & -21.0889 \\
           & 269.6827        & -24.6929 \\
           & 270.8762        & -21.1016 \\

\enddata
\end{deluxetable}


\clearpage
\begin{references}

\reference{}Aptekar, R. et al. 1995, Space Sci. Rev. 71, 265

\reference{}Aptekar, R. et al. 1997, Astron. Lett. 23(2), 147

\reference{}Duncan, R., and Thompson, C. 1992, \apj \, 392, L9

\reference{}Frederiks, D., Aptekar, R., Golenetskii, S., Il'inskii, V., Mazets, E.,
and Terekhov, M., in Gamma-Ray Bursts: 4th Huntsville Symposium, Eds. C. Meegan,
R. Preece, and T. Koshut, AIP Conf. Proc. 428, AIP Press (New York), p. 921, 1998

\reference{}Goss, W., and Shaver, P. 1970, Aust. J. Phys. Astrophysical Suppl. No.
14, 1

\reference{}Gotthelf, E., and Vasisht, G. 1998, New Astronomy 3, 293

\reference{}Green D. A. 1998, A Catalogue of Galactic Supernova Remnants (1998 September version), Mullard Radio Astronomy Observatory, Cambridge, United Kingdom (available at http://www.mrao.cam.ac.uk/surveys/snrs/)

\reference{}Hurley, K. et al. 1992, A\&AS, 92(2), 401

\reference{}Hurley, K. et al. 1999a, \nat \, 397, 41

\reference{}Hurley, K., Kouveliotou, C., Cline, T., Mazets, E., Golenetskii, S., 
Frederiks, D., and van Paradijs, J. 1999b, \apj, submitted

\reference{}Hurley, K. Kouveliotou, C., Woods, P., Cline, T., Butterworth, P., 
Mazets, E., Golenetskii, S., and Frederics, D. 1999c, \apj \, 510, L107 

\reference{}Hurley, K. et al. 1999d, \apj \, 510, L111

\reference{}Hurley, K. et al. 1994, \apj \,423, 709

\reference{}Hurley, K., Kouveliotou, C., Cline, T., and Mazets, E. 1997, \iaucirc \, 6743

\reference{}Kaspi, V., Lyne, A., Manchester, R., Johnston, S., D'Amico, N., and Shemar,
S., \apj \, 409, L57

\reference{}Kouveliotou, C., Norris, J., Wood, K., Cline, T., Dennis, B., Desai, U.,
\& Orwig, L. 1992, \apj \, 392, 179

\reference{}Kouveliotou, C. et al. 1993, \nat \, 362, 728

\reference{}Kouveliotou, C., Fishman, G., Meegan, C., and Woods, P. 1997, \iaucirc \, 6743

\reference{}Kouveliotou, C., et al. 1998, \nat \, 393, 235

\reference{}Kouveliotou, C., et al. 1999, \apj \, 510, L115

\reference{}Li, X., and van den Heuvel, E. 1999, \apj \, 513, L45

\reference{}Marsden, D., Rothschild, R., and Lingenfelter, R. 1999, \apj \,,submitted

\reference{}Mazets, E., Golenetskii, S., and Guryan, Yu. 1979, Sov. Astron. Lett. 5(6), 343

\reference{}Meegan, C. et al. 1996, \apjs \,106, 45

\reference{}Norris, J., Hertz, P., Wood, K., \& Kouveliotou, C. 1991, \apj \,366, 240

\reference{}Reich, W., Reich, P., and F\"{u}rst, E. 1990, Astron. Astrophys. Suppl. Ser.
83, 539

\reference{}Smith, D., Levine, A., Morgan, E., Remillard, R. and Rutledge, R. 1997 
\iaucirc \, 6743

\reference{}Thompson, C., and Duncan, R. 1995, \mnras \, 275, 255

\reference{}White, N., Giommi, P., and Angelini, L. 1995, The WGA Catalog of ROSAT Point Sources, available at  
http://lheawww.gsfc.nasa.gov/users/white/wgacat/wgacat.html

\end{references}
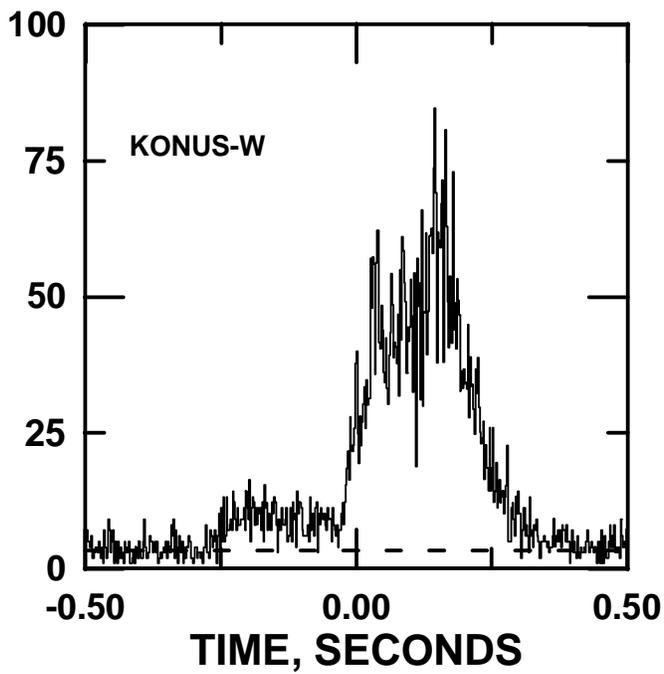
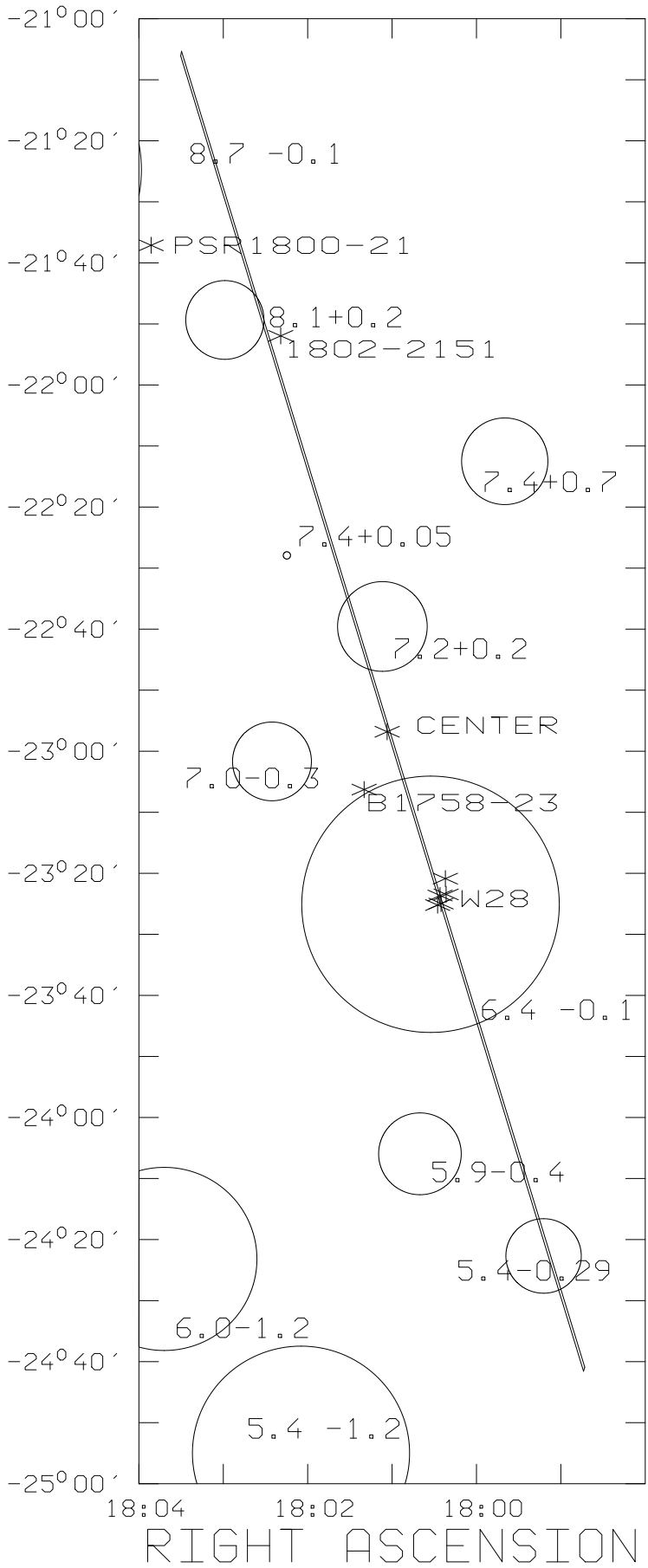
\end{document}